# Magnetic measurements under high pressure with a quantum sensor in Hexagonal Boron Nitride

Lun-Xuan Yu[1,2], Nai-Jie Guo[2,3,4,5], Lin Liu[1,2], Wei Liu[2,3,4,5], Gui-Zhen Yan[1,2], Jin-Ming Cui[2,3,4,5], Jian-Shun Tang[2,3,4,5], Chuan-Feng Li[2,3,4,5], Xiao-Di Liu[1,*]

*[1]Key Laboratory of Materials Physics, Institute of Solid State Physics, HFIPS, Chinese Academy of Sciences, Hefei 230031, China*

*[2]CAS Key Laboratory of Quantum Information, University of Science and Technology of China, Hefei, Anhui 230026, China*

*[3]Anhui Province Key Laboratory of Quantum Network, University of Science and Technology of China*

*[4]CAS Center For Excellence in Quantum Information and Quantum Physics, University of Science and Technology of China, Hefei, Anhui 230026, China*

*[5]Hefei National Laboratory, University of Science and Technology of China, Hefei 230088, China.*

## ABSTRACT

Magnetic measurements under high-pressure conditions are pivotal for the study of superconductivity and magnetic materials but remain challenging due to the micrometer-sized sample in diamond anvil cells (DAC). In this study, we propose a quantum sensing approach utilizing negatively charged boron-vacancy ($V_B^-$) spin defects in two-dimensional hexagonal boron nitride for high resolution magnetic measurements under pressure. The optical and spin properties of $V_B^-$ defects were systematically studied under high-pressure conditions, revealing a significant pressure-induced shift in zero-field splitting (ZFS), approximately three times larger than that of nitrogen-vacancy (NV) center. Furthermore, we demonstrate the pressure-dependent magnetic transition and variations in the Curie temperature of van der Waals ferromagnet $Fe_3GeTe_2$ flake using $V_B^-$ defects under pressures. Notably, the maximum operational pressure for $V_B^-$ defects was determined to be approximately 11 GPa, attributed to a structural phase transition in hexagonal boron nitride (hBN). This work establishes the way for two-dimensional quantum sensing technologies under high-pressure environments.

## INTRODUCTION

Spin defects in wide-bandgap semiconductors, particularly those found in diamond[1,2] and silicon carbide[3,4], have attracted considerable interest owing to their potential in quantum information[5,6], and quantum sensing[7,8]. Among these, the nitrogen-vacancy (NV) centers is particularly notable, offering long spin coherence time[9,10] even at room temperature, which is crucial for quantum technologies. Furthermore, the capability of optically reading and manipulating spin states in these materials enhances their suitability for quantum operations. Despite their advantages, these spin defects face significant limitations[11-13] in practical applications, particularly when utilized as nanoscale sensors. For instance, the dangling bonds[14] on the surface of three-dimensional (3D) substantially degrade spin coherence and reduce sensitivity, posing challenges to their effective deployment.

Against this backdrop, boron-vacancy ($V_B^-$) spin defects in hexagonal boron nitride (hBN) have emerged as highly promising alternatives for quantum sensing applications, addressing many of the limitations associated with traditional spin defects in 3D materials. hBN is a two-dimensional (2D) layered material analogous to graphene, consisting of alternating boron and nitrogen atoms arranged in a honeycomb lattice and bonded via sp² hybridization. Its unique 2D structure allows for excellent interaction between spin defects and the external environment, making it an ideal candidate for integration into 2D nanoscale sensors and other quantum devices. $V_B^-$ defects in hBN are created when a boron atom is removed from the lattice, creating a vacancy that traps an electron. $V_B^-$ defects exhibit spin properties that are sensitive to external perturbations[15-17], enabling precise nanoscale measurements. Despite their promising potential, the structure and behavior of $V_B^-$ defects under extreme conditions, such as high pressure, remains poorly understood.

hBN has been extensively utilized for magnetic measurements under ambient pressure, with its 2D layered structure providing an excellent platform for investigating other 2D materials[18]. Additionally, under high-pressure conditions, NV centers in diamonds and color centers in silicon carbide (SiC) have been successfully deployed to magnetic measurements. These systems have facilitated the study of critical physical phenomena, including the Meissner effect of superconductors[19,20] and magnetic phase transitions[21] in magnetic materials at high pressures. Such studies highlight the indispensable role of quantum sensors in unraveling the complex behaviors of materials under extreme conditions. Despite these advancements, the potential of hBN to perform magnetic measurements under high-pressure conditions remains largely unexplored. Given its inherent 2D nature and the demonstrated sensitivity of $V_B^-$ defects, hBN presents a compelling candidate for achieving high-resolution, two-dimensional nanoscale imaging under elevated pressures. Further research is urgently needed to elucidate the interplay between hBN's structural stability, defect dynamics, and quantum sensing capabilities in such environments, paving the way for breakthroughs in high-pressure quantum sensing technologies.

In this letter, we systematically investigate the pressure-dependent optical and spin properties of $V_B^-$ defects in hBN using fluorescence spectra, optically detected magnetic resonance (ODMR) spectroscopy and Raman spectra under high-pressure conditions. The influence of pressure on ODMR spectra is analyzed, revealing a pronounced pressure-induced shift in the zero-field splitting (ZFS). Notably, this shift is approximately three times larger than that observed in NV centers, highlighting the superior sensitivity of $V_B^-$ defects to pressure variations and their potential for high-precision pressure sensing. Furthermore, we demonstrate the capability of $V_B^-$ defects to probe pressure-dependent magnetic transitions and variations in the Curie temperature of van der Waals ferromagnetic $Fe_3GeTe_2$ flakes. These results underscore the versatility of $V_B^-$ defects as quantum sensors for studying magnetic properties under extreme conditions. Importantly, the maximum operational pressure for $V_B^-$ defects is determined to be approximately 11 GPa, beyond which the structural integrity of hBN is compromised due to a pressure-induced phase transition. This study offers valuable insights into the fundamental behavior of $V_B^-$ defects under extreme conditions, paving the way for their integration into advanced quantum sensing technologies in high-pressure environments.

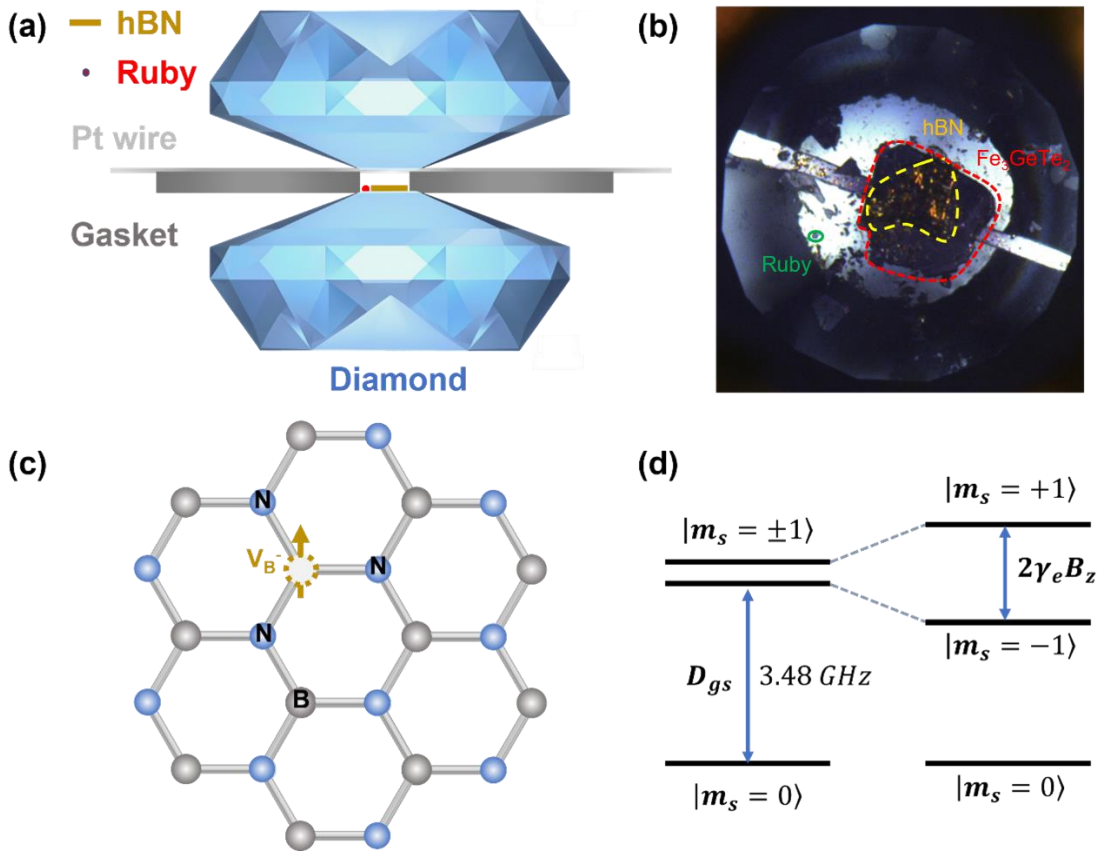

**FIG. 1. The V$_B^-$ defect in hBN in diamond anvil cell (DAC) under high pressure.** (a) Diagram of diamond anvil cell (DAC). A perforated gasket is placed between the two diamonds to form the sample chamber, containing the hBN sample, ruby ball (pressure marker), NaCl (pressure-transmitting medium), and platinum wires for microwave transmission. (b) Optical microscopic image of the sample chamber inside the DAC, showing the position of sample, ruby and platinum wire. (c) Atomic structure of the V$_B^-$ defect in hBN, illustrating the vacancy where a boron atom is missing from the lattice (gray sphere), and the nitrogen atom (blue sphere) that remains in the lattice. (d) Energy level diagram for the V$_B^-$ defect in hBN, showing the spin states and ZFS parameters, as well as the effects of an applied magnetic field.

To investigate the pressure-dependent properties of V$_B^-$ defects in hBN, we employed a diamond anvil cell (DAC) to generate high pressures. The DAC setup, illustrated in Figure 1(a), consists of two diamonds with a perforated gasket between them, forming a small sample chamber. Inside this chamber, we placed the hBN sample along with a ruby ball[22], which served as a pressure marker to calibrate the applied pressure. The sample chamber also contained NaCl as a pressure-transmitting medium and platinum wires to transmit microwaves for ODMR measurements. The optical image of the sample chamber is shown in Figure 1(b). The pressure applied to the sample was controlled by adjusting the distance between the diamonds, and the ruby fluorescence method was used to calibrate the pressure inside the chamber. As pressure increased, the fluorescence emitted by the ruby ball shifted, providing a precise measurement of the applied pressure.

Figure 1(c) shows the atomic structure of $V_B^-$ defects, where a missing boron atom (gray sphere) in the lattice is shown alongside the nitrogen atom (blue sphere), highlighting the vacancy and its electron-trapping nature. The electronic ground state of a $V_B^-$ defects has spin $S = 1$, with its energy level structure illustrated in Figure 1(d). An applied magnetic field induces Zeeman splitting. The ground-state Hamiltonian H[23,24] for negatively charged boron vacancy ($V_B^-$) defects in hBN, accounting for zero-field splitting and interactions with an external magnetic field and applied pressure, is given by:

$$H = D\left(S_z^2 - \frac{1}{3}S(S+1)\right) + E\left(S_x^2 + S_y^2\right) + g\mu_B \boldsymbol{B} \cdot \boldsymbol{S} + [\alpha(\sigma_{xx} + \sigma_{yy}) + \beta\sigma_{zz}]S_z^2 \qquad (1)$$

Where $\mu_B$ is the Bohr magneton, $\boldsymbol{B}$ is the external magnetic field vector, g is the Landé g-factor, which describes the coupling between the magnetic field and the spin, $\boldsymbol{S}$ is the spin operator for the electron spin, $D$ and $E$ are the ZFS parameters. $D$ describes the splitting of the spin states in the absence of an external magnetic field, $E$ is the transverse zero-field splitting parameter, accounting for the anisotropy in the spin states, $S_x$, $S_y$ and $S_z$ are the spin operators in the x, y, and z directions, respectively. Under hydrostatic pressure, $\sigma_{xx}$, $\sigma_{yy}$ and $\sigma_{zz}$ are the diagonal components of the stress tensor $\boldsymbol{\sigma}$, with the off-diagonal terms being zero. These diagonal components should be equal to pressure $P$. Moreover, because hBN is a two-dimensional material, $\beta$ can be neglected compared to $\alpha$.

For the optical excitation of the $V_B^-$ defects, a 532 nm laser with a power of 3.4 mW is used to excite the sample to emit fluorescence. The fluorescence counts are recorded by the single photon counting module. The ODMR spectroscopy technique was employed to measure the spin properties of the $V_B^-$ defects. ODMR is a technique that combines optical excitation and magnetic resonance methods. As laser shining on $V_B^-$ defects to excite electrons, and then applying microwave and magnetic fields to induce magnetic resonance transitions of electrons, the changes in fluorescence intensity caused by these transitions are detected to obtain information about the spin and magnetic properties of $V_B^-$ defects. The spin properties of the $V_B^-$ defects, including spectral shifts and ZFS parameters, were demonstrated at various pressures, ranging from ambient pressure up to 3.2 GPa. The zero-field splitting parameter $D$ was extracted from the spectra, and its dependence on pressure was analyzed to assess the sensitivity of the defects to changes in the pressure environment. These measurements provided insights into how the spin and optical properties of the $V_B^-$ defects evolve under high-pressure conditions.

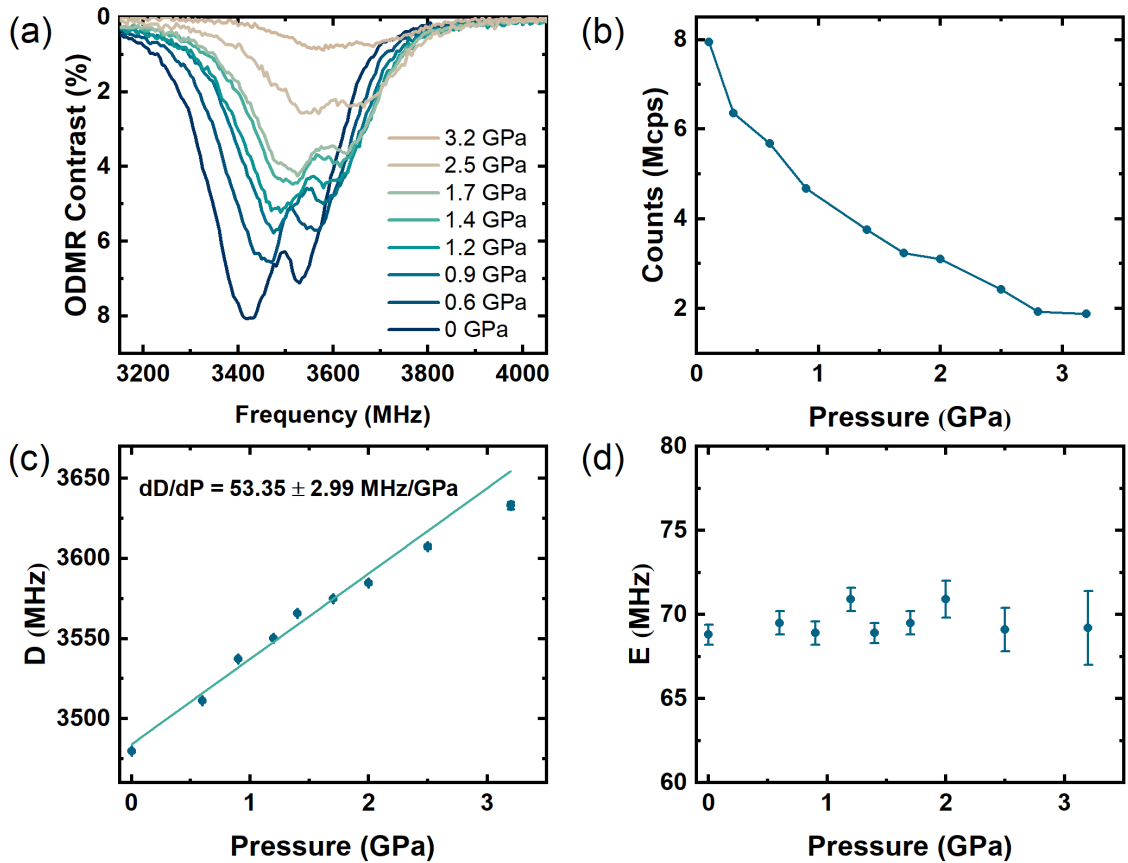

FIG. 2. **Spin properties of $V_B^-$ defects in hBN under compression at room temperature.** (a) Evolution of ODMR spectra for $V_B^-$ defects as pressure increasing. Spectral shifts towards higher frequencies are observed as pressure increased. (b) Fluorescence counts of $V_B^-$ defects at different pressures. A noticeable decrease in fluorescence intensity is shown as pressure increases, disappearing above 3.2 GPa. (c) Zero-field splitting parameter $D$ as a function of pressure, showing a linear dependence with a slope of $53.35 \pm 2.99$ MHz/GPa. (d) Zero-field splitting parameter E as a function of pressure. The parameter E remains nearly constant across the pressure range.

Upon applying pressure to the $V_B^-$ defects in hBN, we observed significant changes in the ODMR spectra. As pressure increased, the spectra shifted towards higher frequencies, as shown in Figure 2(a). This shift was especially pronounced up to 3.2 GPa, above which the ODMR signals began to diminish. The fluorescence intensity, measured simultaneously with the ODMR response, also exhibited a pressure dependence. As depicted in Figure 2(b), the fluorescence counts decreased with increasing pressure, and the signal disappeared entirely above 4.5 GPa, indicating that the optical properties of the $V_B^-$ defects were strongly affected by high pressure.

**TABLE** Ⅰ. Four spin defects in comparison: slope of $D$-$P$ curve, resolution $\eta(P)$ and operational limits $P_{max}$ at room temperature (T = 295 K).

| Spin Defects | dD/dP (MHz/GPa) | $\eta(P)$ (MPa/√Hz) | $P_{max}$ (GPa) |
|---|---|---|---|
| NV Centers[25] | 14.58 | 0.6 | >100 |
| Divacancy Defects[21] | 25.1 | 0.28 | 42 |
| Silicon Vacancy Defects[20] | 0.31 | 0.5 | 27 |
| $V_B^-$ Defects in hBN | 53.35 | 0.24 | 3.2 |

We also studied ZFS parameters, specifically the $D$ and $E$ values, to quantify the pressure dependence of the spin properties. The parameters $D$ and $E$ can be directly extracted from the ODMR spectrum using the following relationships: $D = \frac{v_1 + v_2}{2}$, $E = \frac{v_2 - v_1}{2}$, where $v_1$ and $v_2$ are the observed resonance frequencies. The $D$ value, which describes the splitting between the $m_s = 0$ and $\pm 1$ spin sublevels, exhibited a linear dependence on pressure [Figure 2(c)]. The slope of this relationship was determined to be $53.35 \pm 2.99$ MHz/GPa, which is around three times larger than that of NV centers in diamond, indicating the promising application as pressure quantum sensor.

With pulsed-ODMR spectra, pressure sensitivity $\eta$(P) of 0.24 MPa•Hz$^{-1/2}$ is obtained (see supplemental information). The pressure sensitivity of $V_B^-$ defects surpasses that of other spin defects, albeit with a lower maximum operational pressure $P_{max}$ of 3.2 GPa than other spin defects, as shown in Table 1. In contrast, the $E$ value, which represents the transverse splitting, remained nearly constant across the pressure range, as shown in Figure 2(d). This indicates that the pressure-induced changes predominantly affect the longitudinal splitting (represented by $D$), while the transverse splitting (represented by $E$) remains largely unaffected.

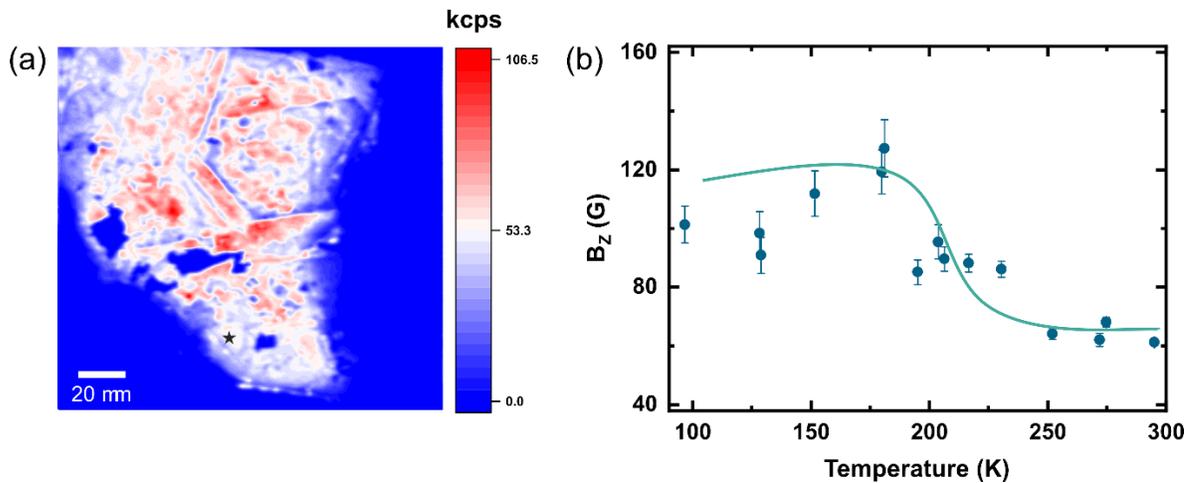

**FIG. 3. The measurements of the magnetic phase transition of $Fe_3GeTe_2$ using $V_B^-$ defects.** (a) Confocal scanning microscopy image of $V_B^-$ defects with $Fe_3GeTe_2$ sample below. The black star is the detected position. (b) The temperature-dependent changes in the magnetic field magnitude perpendicular to the surface of $Fe_3GeTe_2$, measured using $V_B^-$ defects.

Based on the optical and spin properties studied under high-pressure, we extended our investigation to explore magnetic measurements of the van der Waals ferromagnet $Fe_3GeTe_2$ flakes using $V_B^-$ defects in hBN. Among the cleavable, layered 2D van der Waals magnetic materials, $Fe_3GeTe_2$ is particularly notable due to its unique properties, including itinerant ferromagnetism (FM) and a relatively high Curie temperature ($T_c$) of approximately 220 K in its bulk form[26]. These intrinsic attributes, combined with the strong interfacial compatibility between $Fe_3GeTe_2$ and hBN, make it an exceptional candidate for quantum sensing applications under high-pressure conditions. Below its pressure-sensitive Curie temperature, the magnetization of $Fe_3GeTe_2$ aligns perpendicularly to its surface, coinciding with the spin orientation of $V_B^-$ defects in hBN. This alignment enables precise measurements of the magnetic field component perpendicular to the material's surface.

Figure 3(a) illustrates the confocal scanning microscopy image of hBN with $Fe_3GeTe_2$ tightly attached beneath it, where the position of detected $V_B^-$ defects is marked by black star. This experimental configuration ensures precise alignment for detecting the local out-of-plane magnetic field perpendicular to the surface of $Fe_3GeTe_2$. The out-of-plane magnetic field component $B_z$, shown in Figure 3(b), was experimentally determined by monitoring the ODMR frequency shifts of $V_B^-$ defects. Upon applying microwave and magnetic fields, the Zeeman effect caused linear splitting of the spin sublevels, which are monitored by the microwave frequency resonance peaks in the ODMR spectrum. By analyzing the linear splitting of resonance peak, $B_z$ was quantitatively calculated using the relation: $E = g\mu_B B_z/\hbar$. In this study, the $E$ value of $V_B^-$ defects was leveraged to detect magnetic field variations during the phase transition of $Fe_3GeTe_2$. We demonstrated the magnetic phase transition of $Fe_3GeTe_2$ from a paramagnetic to a ferromagnetic state, and the Curie temperature is about 210 K at under pressure 0.7 GPa, which is consistence with the results reported previously[26]. The high sensitivity of $V_B^-$ defects enabled detailed observation of subtle magnetic changes, highlighting their effectiveness as quantum sensors for probing the magnetic properties and phase transitions of low-dimensional materials under high-pressure conditions.

Since the ODMR signals of $V_B^-$ defects can only be measured up to about 3.2 GPa now, an open question remains as what's the maximum operational pressure of $V_B^-$ defects. Here we explore the pressure limit of $V_B^-$ defects. As shown in Figure 2(a), at pressures above 3.2 GPa, the ODMR signals vanished, and we observed a dramatic change in the optical properties of the sample. The Raman spectra, shown in Figure 4(a), revealed a shift of the characteristic $E_{2g}$ phonon mode to higher wavenumbers with increasing pressure, which agrees with previous studies[27]. Above 10.7 GPa, the $E_{2g}$ mode completely disappeared, suggesting the phase transition of the hexagonal phase of BN into another phase. This observation consistence with previous studies indicating that hBN undergoes phase transitions at high pressures. Simultaneously, the photoluminescence (PL) spectra of $V_B^-$ defects displayed a striking pressure-dependent behavior, as illustrated in Figure 4(b). While the PL intensity gradually decreased with increasing pressure, a sudden and complete disappearance of the PL signal was observed beyond 4.5 GPa. This abrupt loss suggests that the electronic states associated with $V_B^-$ defects are highly sensitive to lattice distortions and structural in hBN. The structural instability of single-crystal hBN is important for the high-pressure sensing applications. Therefore, we identify a maximum operational pressure of approximately 11 GPa for $V_B^-$ defects in single-crystal hBN, which was explained and evidenced by a structural phase transition in hBN.

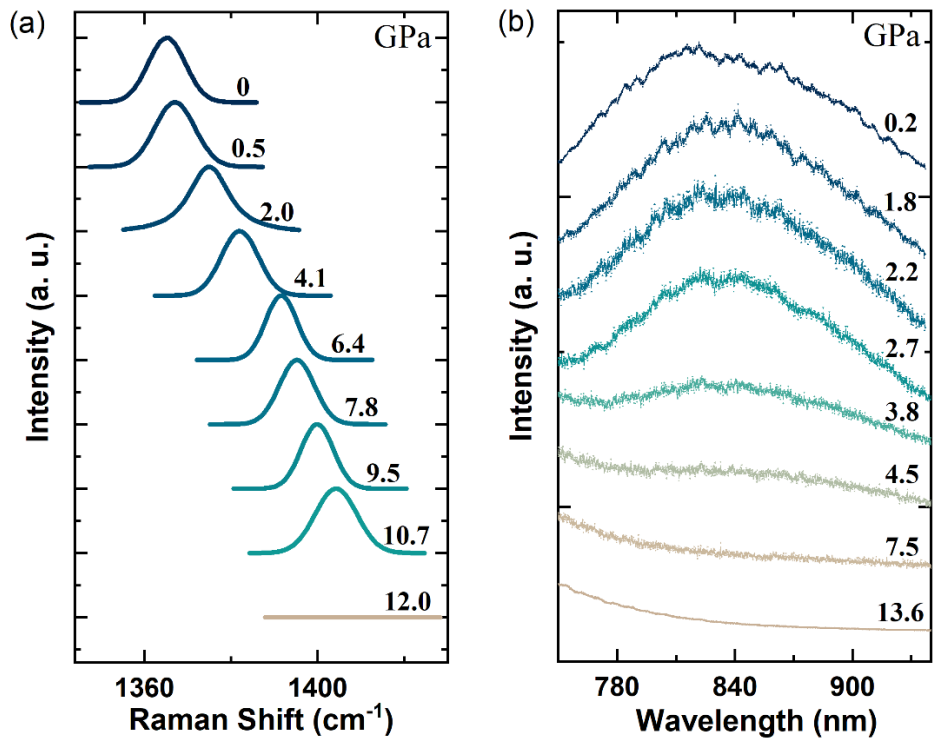

**FIG. 4. the Raman spectra of hBN and the photoluminescence spectra of $V_B^-$ defects under varying pressures.** (a) Raman spectra of hBN at different applied pressures, showing the shift of the characteristic $E_{2g}$ phonon mode to higher wavenumbers as pressure increases. Above 10.7 GPa, the mode completely disappears, suggesting the loss of the hexagonal phase of BN. (b) Photoluminescence spectra of $V_B^-$ defects under varying pressures, above 4.5 GPa, the photoluminescence spectra of $V_B^-$ defects completely disappear.

This study explores the pressure-dependent properties of $V_B^-$ defects in hBN, demonstrating their potential for quantum sensing applications. We observed that the ZFS parameter $D$ exhibits a linear dependence on pressure, yielding a high pressure sensitivity of $\eta(P)$ of 0.24 MPa•Hz$^{-1/2}$, surpassing that of other spin defects such as NV centers in diamond. In contrast, the transverse splitting parameter E remains unaffected by pressure, indicating that pressure-induced changes predominantly impact the longitudinal spin states. Beyond 3.2 GPa, ODMR signals vanish, and above 10.7 GPa, a phase transformation occurs in hBN. In addition to these fundamental insights, we also demonstrated the ability of $V_B^-$ defects to serve as high-resolution magnetic field sensors, specifically for detecting the magnetic phase transition of the van der Waals ferromagnet $Fe_3GeTe_2$. The pressure-sensitive nature of $V_B^-$ defects allows for precise measurement of the magnetic transition in $Fe_3GeTe_2$, showcasing their utility for studying the magnetic properties of low-dimensional materials. These findings highlight the potential of $V_B^-$ defects as quantum sensors for magnetic measurements under high pressure, a critical capability for investigating phase transitions in superconductors and magnetic materials. Ultimately, this work establishes $V_B$ defects in hBN as promising tools for high-pressure quantum sensing, opening up new possibilities for studying exotic magnetic phases and advancing quantum technologies. Future research could further explore the integration of these defects into quantum sensors, expanding their applications in a wide range of scientific and technological fields.

*Acknowledgements*-This work was supported by research grants of the Youth Innovation Promotion Association of CAS (No. 2021446), the National Science Foundation of China (Grants No. 12204484, No. 11874361), Anhui key research and development program(2022h11020007), and the HFIPS Director's Fund of Chinese Academy of Sciences (Nos. BJPY2023B02).

* To whom correspondence should be addressed: xiaodi@issp.ac.cn

## REFERENCENS